\documentclass[twocolumn, superscriptaddress]{aastex63}
\usepackage{CJK}
\usepackage{url}
\usepackage{amsmath}
\usepackage[super]{nth}
\usepackage{microtype}
\usepackage{graphicx}

\submitjournal{AAS Journals}

\shorttitle{KELT-9 b for the win(d)s}
\shortauthors{Pai Asnodkar et al.}

\begin{document}
\begin{CJK*}{UTF8}{gbsn}

\title{Variable and super-sonic winds in the atmosphere of an ultra-hot giant planet}

\author[0000-0002-8823-8237]{Anusha Pai Asnodkar}
\affiliation{The Ohio State University, McPherson Laboratory, 140 W 18th Ave., Columbus, OH 43210, USA}
\author[0000-0002-4361-8885]{Ji Wang (王吉)}
\affiliation{The Ohio State University, McPherson Laboratory, 140 W 18th Ave., Columbus, OH 43210, USA}
\author[0000-0003-3773-5142]{Jason D.\ Eastman}
\affiliation{Center for Astrophysics \textbar \ Harvard \& Smithsonian, 60 Garden St, Cambridge, MA 02138, USA}
\author[0000-0001-9207-0564]{P. Wilson Cauley}
\affiliation{Laboratory for Atmospheric and Space Physics, University of Colorado Boulder, Boulder, CO 80303}
\author{B. Scott Gaudi}
\affiliation{The Ohio State University, McPherson Laboratory, 140 W 18th Ave., Columbus, OH 43210, USA}
\author{Ilya Ilyin}
\affiliation{Leibniz-Institute for Astrophysics Potsdam (AIP), An der Sternwarte 16, D–14482 Potsdam, Germany}
\author[0000-0002-6192-6494]{Klaus Strassmeier}
\affiliation{Leibniz-Institute for Astrophysics Potsdam (AIP), An der Sternwarte 16, D–14482 Potsdam, Germany}

\correspondingauthor{Anusha Pai Asnodkar}
\email{paiasnodkar.1@osu.edu}

\begin{abstract}
Hot Jupiters receive intense irradiation from their stellar hosts. The resulting extreme environments in their atmospheres allow us to study the conditions that drive planetary atmospheric dynamics, e.g., global-scale winds. General circulation models predict day-to-nightside winds and equatorial jets with speeds on the order of a few km $\mathrm{s^{-1}}$. To test these models, we apply high-resolution transmission spectroscopy using the PEPSI spectrograph on the Large Binocular Telescope to study the atmosphere of KELT-9 b, an ultra-hot Jupiter and currently the hottest known planet. We measure $\sim$10 km $\mathrm{s^{-1}}$ day-to-nightside winds traced by Fe II features in the planet's atmosphere. This is at odds with previous literature (including data taken with PEPSI), which report no significant day-to-nightside winds on KELT-9 b. We identify the cause of this discrepancy as due to an inaccurate ephemeris for KELT-9 b in previous literature. We update the ephemeris, which shifts the mid-transit time by up to 10 minutes for previous datasets, resulting in consistent detections of blueshifts in all the datasets analyzed here. Furthermore, a comparison with archival HARPS-N datasets suggests temporal wind variability $\sim$5-8 km $\mathrm{s^{-1}}$ over timescales between weeks to years. Temporal variability of atmospheric dynamics on hot Jupiters is a phenomenon anticipated by certain general circulation models that has not been observed over these timescales until now. However, such large variability as we measure on KELT-9 b challenges general circulation models, which predict much lower amplitudes of wind variability over timescales between days to weeks. 
\end{abstract}


\section{Introduction} \label{sec:intro}
Among transiting exoplanets, hot Jupiters (HJs) are the most accessible and exotic class of planets for atmospheric characterization. Their large size and proximity to their host stars establishes them both as ideal targets for transit techniques and unique laboratories for exploring the interplay between radiative, chemical, and dynamical processes in gas giant atmospheres. Transmission and emission spectroscopy studies suggest that more highly irradiated HJ atmospheres display a diversity of thermal structures. 

\begin{table*}[t]
\caption{Datasets used to study the atmospheric dynamics of KELT-9 b and KELT-20 b: columns provide the name of the instrument, the date of the starting observation, the start and end times of observing, and the number of observations.}
\centering
    \begin{tabular}{llllll}
    \hline
    \hline
    Target & Instrument & Night & $t_{\mathrm{start}}$ (UTC) & $t_{\mathrm{end}}$ (UTC) & $N_{\mathrm{obs}}$ \\ 
    \hline
    KELT-9 b & HARPS-N & 2017-07-31 & 20:59:04.0 & 05:19:10.0 & 49 \\
    & PEPSI & 2018-07-03 & 04:07:34.4 & 11:16:41.3 & 82 \\
    & HARPS-N & 2018-07-20 & 21:20:24.0 & 05:09:58.0 & 46 \\ 
    & PEPSI & 2019-06-22 & 05:19:33.7 & 11:29:06.5 & 65 \\
    \hline
    KELT-20 b & HARPS-N & 2017-08-16 & 22:26:04.7 & 04:00:39.4 & 90 \\
    & HARPS-N & 2018-07-12 & 21:27:47.4 & 05:14:51.9 & 116 \\
    & HARPS-N & 2018-07-19 & 21:26:28.1 & 04:24:36.4 & 78 \\
    & PEPSI & 2019-05-04 & 07:23:06.0 & 11:57:55.7 & 22 \\
    \hline
    \end{tabular}
\label{tab:datasets}
\end{table*}

\par Recent efforts in exoplanet characterization seek empirical constraints on HJ atmospheric circulation \citep{Snellen2010, Ehrenreich2020, Cauley2021}. Atmospheric dynamics are a critical form of energy transport and significantly contribute to the full energy budget of highly irradiated exoplanet atmospheres. The strong temperature contrast between the day and nightsides of hot Jupiters drives global day-to-nightside winds and superrotating equatorial jets. General circulation models (GCMs) of classic HJs like HD 209458 b or HD 189733 b predict strong ($\sim$1-3 km $\mathrm{s^{-1}}$) day-to-nightside winds \citep{Showman2013}. Equatorial jets ($\sim$3-4 km $\mathrm{s^{-1}}$) are predicted to offset the dayside hotspot eastward of the sub-stellar point \citep{Showman2008, Showman2011}. Notably, few GCMs that explore low pressure dynamics exist, especially in the millibar to $\mu$bar range relevant to high-resolution transmission spectroscopy observations \citep{Hoeijmakers2019}. 

\par GCM predictions for HJs have been validated by phase curve observations \citep{Knutson2007, Kataria2016, Wong2020} and high-resolution Doppler spectroscopy \citep{Snellen2010, Louden2015}. Global day-to-nightside winds can be probed with transmission spectroscopy between the 2nd and 3rd contact of a transit to measure the terminator-averaged spectrum of the planet's atmosphere. During transit, day-to-nightside winds manifest as lines that are blueshifted relative to the planet's orbital velocity. Empirical measurements of these winds on HJs thus far corroborate the $\sim$few km $\mathrm{s^{-1}}$ prediction from GCMs. The first such measurement was traced by CO absorption in the near-infrared, revealing $2 \pm 1$ km $\mathrm{s^{-1}}$ day-to-nightside winds on HD 209458 b \citep{Snellen2010}. A study of CO and $\rm{H_2O}$ absorption features retrieved day-to-nightside wind speeds of $1.7^{+1.1}_{-1.2}$ km $\mathrm{s^{-1}}$ on HD 189733 b \citep{Brogi2016}. Wind velocity and hotspot offset measurements provide direct evidence for the extremity and relevance of atmospheric dynamics in shaping the thermal structure of highly irradiated giant planets. Atmospheric dynamics in HJs can also have first-order effects on the template spectra that should be used in the analysis and interpretation of high-resolution spectroscopy observations \citep{Beltz2021a, Flowers2019}. 

\par In this work, we report on the atmospheric dynamics of KELT-9 b \citep{Gaudi2017} and KELT-20 b \citep{Lund2017, Talens2018}, both of which fall in a categorically distinct class of objects known as ultra-hot Jupiters (UHJs). The unique atmospheric thermal structures and chemical composition of giant planets with $T_{\mathrm{eq}} \gtrsim 2500$ K motivate this separate class of HJs. Most notably, thermal inversions (regions of the atmosphere analogous to Earth's stratosphere where temperature increases with decreasing pressure, i.e. increasing altitude) only occur in the hottest HJs. The majority of ultra-hot Jupiters display nearly isothermal pressure-temperature (P-T) profiles around milibar pressures and lower \citep{Lothringer2019} while some decrease outward in temperature \citep{Madhusudhan2019}. A small subset of UHJs exhibit evidence of thermal inversions, including the classic examples WASP-18 b \citep{Sheppard2017}, WASP-121 b \citep{Evans2017}, WASP-33 b \citep{Haynes2015}, and our target of focus, KELT-9 b \citep{Pino2020}. 

\begin{table*}[t]
\caption{KELT-9 system parameters (reproduced from \citealt{PaiAsnodkar2021}).}
\centering
    \begin{tabular}{lllll}
    \hline
    \hline
    Parameter & Units & Symbol & Value & Source \\ 
    \hline
    Stellar parameters: \\
    \hspace{3mm} Stellar mass & $M_\odot$ & $M_\star$ & $2.52^{+0.25}_{-0.20}$ & \citealt{Gaudi2017} \\
    \hspace{3mm} Stellar radius & $R_\odot$ & $R_\star$ & $2.362^{+0.075}_{-0.063}$ & \citealt{Gaudi2017}\\
    \hspace{3mm} Stellar density & g $\mathrm{cm^{-3}}$ & $\rho_\star$ & $0.2702 \pm 0.0029$ & \citealt{Gaudi2017} \\
    \hspace{3mm} Effective temperature & K & $T_{\mathrm{eff}}$ & $10170 \pm 450$ & \citealt{Gaudi2017}\\
    \hspace{3mm} Projected rotational velocity & km s$^{-1}$ & $v\sin{i}$ & $111.4 \pm 1.3$ & \citealt{Gaudi2017} \\
    Planetary parameters: \\
    \hspace{3mm} Planet mass & $M_J$ & $m_{\mathrm{p}}$ & $2.88 \pm 0.84$ & \citealt{Gaudi2017} \\
    \hspace{3mm} Planet radius & $R_J$ & $R_{\mathrm{p}}$ & $1.891^{+0.061}_{-0.053}$ & \citealt{Gaudi2017} \\
    \hspace{3mm} Semi-major axis & AU & $a$ & $0.03462^{+0.00110}_{-0.00093}$ & \citealt{Gaudi2017} \\
    \hspace{3mm} Eccentricity & & $\varepsilon$ & $0$ & \citealt{Gaudi2017} \\
    \hspace{3mm} Spin-orbit alignment & degree & $\lambda$ & $-84.8\pm1.4$ & \citealt{Gaudi2017} \\
    \hspace{3mm} Orbital inclination & degree & $i_{\mathrm{orbit}}$ & $86.79\pm0.25$ & \citealt{Gaudi2017} \\
    Ephemeris: \\
    \hspace{3mm} Mid-transit time & $\mathrm{BJD_{TDB}}$ & $T_0$ & 2458566.436560 $\pm$ 0.000048 & This work \\
    \hspace{3mm} Time of secondary eclipse & $\mathrm{BJD_{TDB}}$ & $T_\mathrm{S}$ & 2458584.950546 $\pm$ 0.000048 & This work \\
    \hspace{3mm} Orbital period & days & $P$ & 1.48111890 $\pm$ 0.00000016 & This work \\
    \hspace{3mm} Ingress/egress transit duration & days & $\tau$ & $0.012808^{+0.000027}_{-0.000026}$ & This work \\
    \hspace{3mm} Total transit duration & days & $T_{14}$ & $0.15949\pm0.00011$ & This work \\
    \hline
    \end{tabular}
\label{tab:system_params}
\end{table*}

\par UHJs are further distinguished by their composition and lack of molecular features (e.g. water, molecular hydrogen) due to dissociation at extreme temperatures. Extending giant planet GCMs to the extreme pressure-temperature regime of UHJs requires introducing additional chemistry due to the dissociation of molecular hydrogen and magnetic effects due to metal ionization. For this reason, UHJs are challenging to model and are relatively new territory in GCM literature. Accounting for the thermodynamic effect of $\rm{H_2}$ dissociation has been shown to horizontally (i.e. across longitudes) homogenize heat redistribution \citep{Roth2021}, resulting in weaker day-to-nightside winds than expected without dissociation \citep{Tan2019}. Wind measurements may answer questions regarding the role of second-order effects in UHJ atmospheres like molecular dissociation and magnetic drag from thermally ionized winds \citep{Perna2010, Rogers2014}. 


\par Previous studies of atmospheric dynamics on KELT-9 b report no significant day-to-nightside winds \citep{Yan2018, Yan2019, Cauley2019, Hoeijmakers2019, Wyttenbach2020}. In this work, we present new day-to-nightside wind measurements of KELT-9 b that challenge previous measurements and theoretical predictions. We employ two epochs of observations from the Potsdam Echelle Polarimetric and Spectrographic Instrument (PEPSI) \citep{Strassmeier2015} on the Large Binocular Telescope (LBT). We apply both a high resolution line-by-line analysis and the cross-correlation technique to study the Doppler signature of Fe II in the atmosphere of KELT-9 b. We compare the PEPSI findings with archival data from the High Accuracy Radial velocity Planet Searcher for the Northern hemisphere (HARPS-N) originally presented in \citealt{Hoeijmakers2019}. We also validate our methods against previous literature using KELT-20 b as a test case. We discuss our wind measurements across techniques and datasets, compare with previous literature, and discuss the consequences for GCMs of UHJs.

\section{Methods} \label{sec:method}

\par
\begin{figure*}[t]
    \includegraphics[width=1\textwidth]{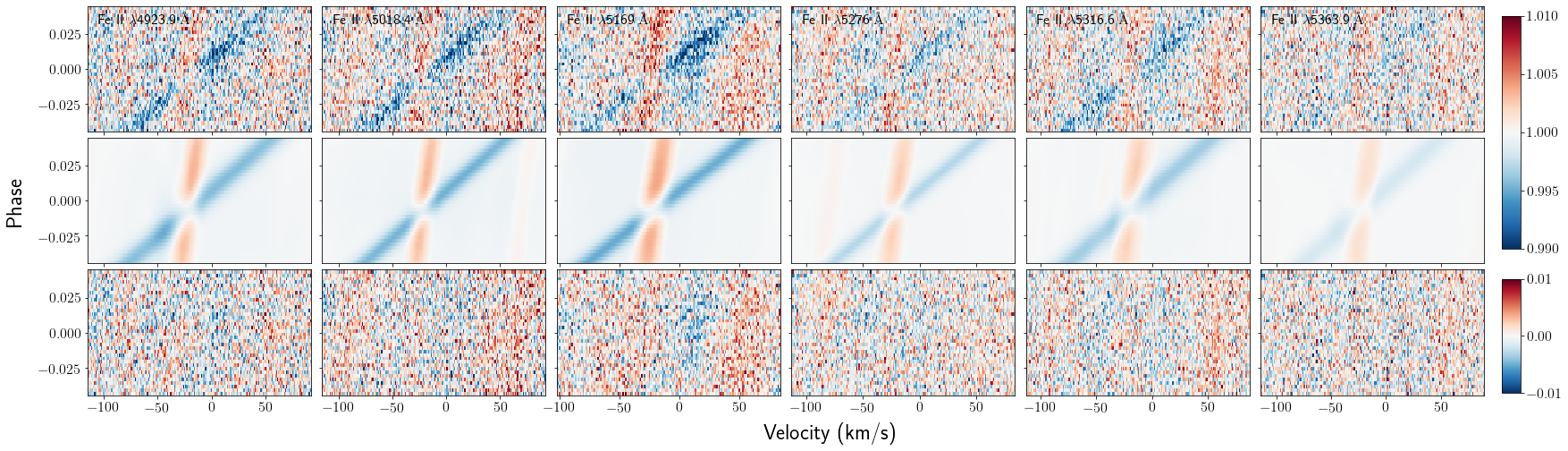}
    \caption{2D maps  of transmission spectra, focusing on the six Fe II absorption lines in the PEPSI 2018 dataset chosen for fitting $K_{\mathrm{p}}$ and $v_{\rm{wind}}$. The blue track is the planet's atmospheric absorption while the red track is the Doppler shadow from the Rossiter-McLaughlin effect; both tracks only form during a transit. Top panel displays fully in-transit observations. Middle panel shows the best-fit model from MCMC sampling. The bottom panel presents the residuals (data - model).}
    \label{fig:FeIIlines}
\end{figure*}

 Our methodology is presented in greater detail in our previous paper \citealt{PaiAsnodkar2021}, but we will summarize the key points here. We compile observations of the KELT-9 system from the high-resolution spectrograph PEPSI (R=50,000) on the LBT (see Figure \ref{fig:FeIIlines}) and archival data from HARPS-N (R=115,000) on Telescopio Nazionale Galileo (TNG). Our analysis of PEPSI data exclusively employs observations from the blue arm with cross-disperser 3 ($\sim$4750--5430 \text{\AA}). For comparison, we select the same wavelength range in our examination of the HARPS-N data. As a test, we also perform a homogeneous analysis of KELT-20 b's atmospheric dynamics traced by the same six Fe II lines and wavelength range for cross-correlation we investigate in KELT-9 b's atmosphere. We analyze an original PEPSI dataset as well as archival HARPS-N observations of the KELT-20 system. We refer to the KELT-9 b datasets by their instrument and year of observation (HARPS-N 2017, PEPSI 2018, HARPS-N 2018, PEPSI 2019, and HARPS-N 2018 in chronological order). To avoid confusion, we refer to the KELT-20 datasets by their instrument and full date of observation (HARPS-N 20170816, HARPS-N 20180712, HARPS-N 20180719, and PEPSI 20190504). Further details about the observations are presented in Table \ref{tab:datasets} and \S 2 of \citealt{PaiAsnodkar2021}. 

\par We employ transmission spectroscopy to probe the atmospheric dynamics of KELT-9 b and KELT-20 b. We apply least-squares deconvolution (\S 3.1 of \citealt{PaiAsnodkar2021}) on out-of-transit observations to determine stellar radial velocities and fit for a circular stellar orbital solution to quantify the systemic velocities for both systems as individually measured by the PEPSI and HARPS-N instruments ($v_{\mathrm{sys, PEPSI}} = -17.86\pm0.044$ km s$^{-1}$ and $v_{\mathrm{sys, HARPS-N}} = -17.15\pm0.11$ km s$^{-1}$)). We divide out the stellar component encapsulated by the out-of-transit observations to recover the transmission spectra and isolate the planet's atmospheric absorption signature (\S 3.2 of \citealt{PaiAsnodkar2021}). We perform both a line-by-line analysis and cross-correlation analysis of the fully in-transit observations to avoid asymmetries and velocity offsets from equatorial jets and rotation during ingress/egress \citep{Cauley2021, Casasayas-Barris2019, Yan2018}. We generate a grid of models of the secondary effects from the Rossiter-McLaughlin effect (RME) and center-to-limb variation (CLV). We simultaneously fit the in-transit secondary effects and planetary atmospheric absorption signatures from individual Fe II lines and cross-correlation. We model the planet's absorption as a Gaussian feature shifted according to a circular orbital solution modified by a constant blueshift ($v_{\rm{wind}}$) due to day-to-nightside winds. See \citealt{PaiAsnodkar2021} for more details.

\par A critical aspect in our analysis of the KELT-9 system is our revised ephemeris. We refit the KELT-9 system with \texttt{EXOFASTv2} \citep{Eastman2019} to include Transiting Exoplanet Survey Satellite (TESS) \citep{TESS2015} lightcurves from 2019 in addition to the follow up lightcurves and TRES RVs from the discovery paper. The updated ephemeris is critical for this analysis since the observations span multiple years and precision deteriorates as the ephemeris is propagated over a longer temporal baseline. In our analysis of KELT-20 b, we adopt the system ephemeris given in \citealt{Lund2017} (the \citealt{Talens2018} ephemeris does not yield significantly different wind velocities). We do not update the KELT-20 ephemeris because the analysis of its atmospheric dynamics is less sensitive to mid-transit timing due to its slower orbital velocity and larger orbital period.

\section{Results} \label{sec:results}

\begin{figure*}[t]
    \begin{minipage}{0.48\linewidth}
        \centering
        \includegraphics[width=\textwidth]{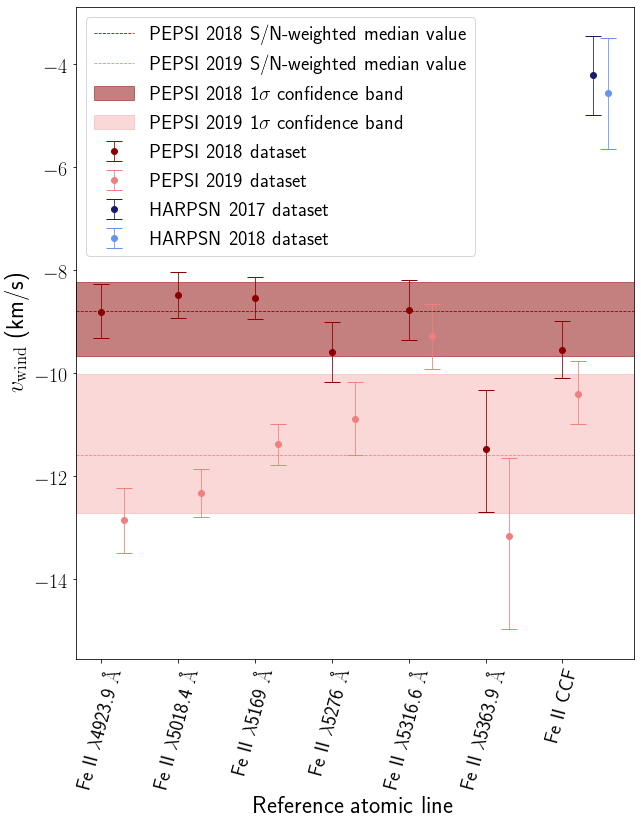} \\
        (a)
    \end{minipage}\hfill
    \begin{minipage}{0.48\linewidth}
        \centering
        \includegraphics[width=\textwidth]{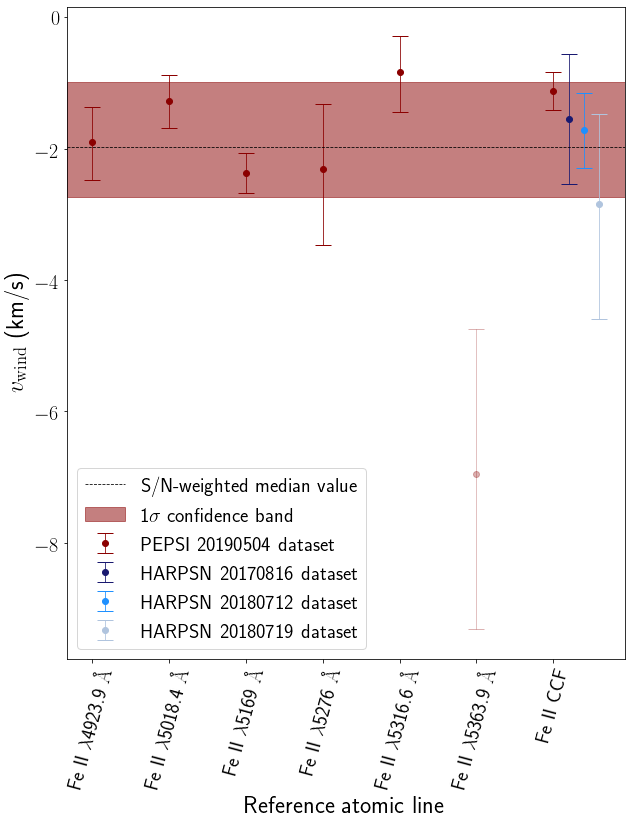}\\
        (b)
    \end{minipage}
    \caption{Day-to-nightside winds measurements from Fe II line-by-line and cross-correlation analyses with MCMC sampling errors for (a) KELT-9 b and (b) KELT-20 b. We note that the Fe II $\lambda$5363.9 {\AA} line is very weak with  S/N = 1.6; for comparison, the Fe II $\lambda$5169 {\AA} line has S/N = 9.0 (see Figure \ref{fig:FeIIlines} for the observed relative line strength of KELT-9 b as an example). Thus, this line is less reliable for measuring day-to-nightside winds.}
    \label{fig:vwind}
\end{figure*}

The left panel of Figure \ref{fig:vwind} summarizes the global day-to-nightside winds observed in KELT-9 b's PEPSI 2018 and PEPSI 2019 datasets. Across the six different Fe II lines we inspected, the measurements are generally consistent within a given dataset. However, there is a systematic offset between the 2018 and 2019 measurements. The error-weighted median value with 1$\sigma$ errors of the 2018 and 2019 measurements across all six lines are $-8.79^{+0.56}_{-0.88}$ km $\mathrm{s^{-1}}$ and $-11.59^{+1.56}_{-1.13}$ km $\mathrm{s^{-1}}$ respectively.

\par We also measure day-to-nightside winds from the two PEPSI datasets as traced by Fe II absorption using cross-correlation. This analysis yields $-9.55^{+0.56}_{-0.55}$ km $\mathrm{s^{-1}}$ and $-10.41^{+0.64}_{-0.57}$ km $\mathrm{s^{-1}}$ for the PEPSI 2018 and PEPSI 2019 datasets respectively. Thus the cross-correlation measurements are consistent with the line-by-line analysis of their corresponding datasets. 

\par In contrast, the HARPS-N measurements taken from cross-correlation differ by a significant $\sim$5-8 km $\mathrm{s^{-1}}$ offset. In particular, the PEPSI 2018 and HARPS-N 2018 datasets were both taken in July and differ by 5.7$\sigma$ (5.3 km $\mathrm{s^{-1}}$). This discrepancy is not attributable to an RV offset between the two instruments since the systemic velocities we obtain for both instruments from our analysis of stellar RVs are consistent within 1 km $\mathrm{s^{-1}}$. As there is no evidence to-date of a companion in the system to cause significant transit-timing variations and upon confirming that our conversion to the $\rm{BJD_{TDB}}$ timing system for all datasets was applied correctly (see Section 2 of \citealt{PaiAsnodkar2021}), we conclude that this discrepancy in our velocity measurements is most plausibly astrophysical. The discrepancy suggests multi-epoch temporal variability of global day-to-nightside winds on KELT-9 b.

\section{Discussion} \label{sec:discussion}

\subsection{Comparison with previous literature}

\begin{figure*}[t]
    \centering
    \includegraphics[width=0.7\textwidth]{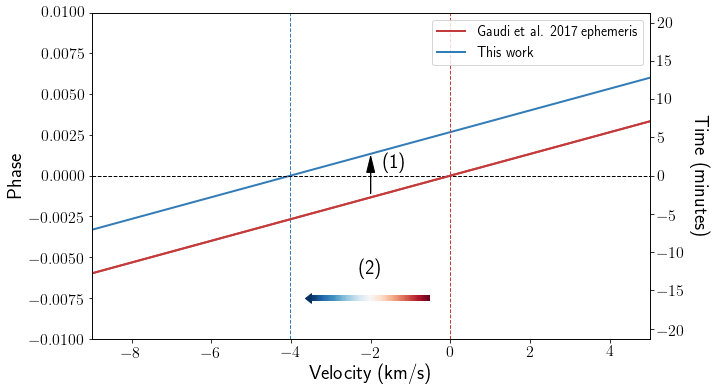}
    \caption{The sloped solid lines represent the center of the planet's absorption signature over the course of the transit, corrected by the systemic velocity (see Figure 2 of \citealt{PaiAsnodkar2021}). The slope captures the orbital motion of the planet while the velocity at mid-transit ($\phi = 0$) corresponds to the day-to-nightside wind velocity. \nth{2} and \nth{3} contact occur at orbital phases beyond the range of this plot. Our revised ephemeris for the KELT-9 system (1) shifts the atmospheric signature up in phase, resulting in (2) a more blueshifted measurement of day-to-nightside winds.}
    \label{fig:KELT9_ephemerisWindShift}
\end{figure*}

\par Using the HARPS-N datasets, our results are blueshifted relative to the measurements in \citealt{Hoeijmakers2019} by $\sim$4 km $\mathrm{s^{-1}}$. They originally measured no significant atmospheric blueshift with these same data (statistically consistent with $v_{\rm{wind}} = 0$ km $\mathrm{s^{-1}}$) as traced by numerous metal species, including Fe II, using cross-correlation. We can account for this discrepancy due to our revised mid-transit time ephemeris; see Table \ref{tab:system_params} (also given as Table 2 in \citealt{PaiAsnodkar2021}). The revised ephemeris results in a propagated mid-transit time that is 4.2 minutes earlier for the HARPS-N 2017 epoch and 5.8 minutes earlier for the HARPS-N 2018 epoch than the original ephemeris presented in \citealt{Gaudi2017} which \citealt{Hoeijmakers2019} adopt. According to our measurement of the orbital motion of the planet, these revised mid-transit times result in day-to-nightside winds that are correspondingly blueshifted by 3.0 km $\mathrm{s^{-1}}$ and 4.1 km $\mathrm{s^{-1}}$ relative to the measurement using the original mid-transit time ephemeris.

\par For a visual understanding of how mid-transit timing affects the measured day-to-nightside winds, refer to Figure \ref{fig:KELT9_ephemerisWindShift}. We define orbital phase as $\phi = \frac{t - T_0}{P}$ such that $\phi=0$ corresponds to mid-transit. At this phase, the planet's atmospheric signature is only shifted by the day-to-nightside winds (given that the data has been corrected for the systemic velocity, the orbital radial velocity of the planet at mid-transit is 0 km $\mathrm{s^{-1}}$). \citealt{Hoeijmakers2019} found no significant day-to-nightside winds, so the velocity of the atmospheric absorption signature they measure (represented by the red solid curve in Figure \ref{fig:KELT9_ephemerisWindShift}) at $\phi=0$ is 0 km $\mathrm{s^{-1}}$. According to our definition of orbital phase, an earlier mid-transit time effectively shifts the atmospheric absorption signature (and the whole flux map) up in phase. Based on the orientation of the atmospheric absorption signature during transit, shifting the absorption signature up in phase according to our revised ephemeris (represented by the blue solid line in Figure \ref{fig:KELT9_ephemerisWindShift}) results in a more blueshifted absorption signature during mid-transit and thus yields a more blueshifted measurement of day-to-nightside winds. 

\par Likewise, \citealt{Cauley2019} reports no significant day-to-nightside winds on KELT-9b by studying individual metal lines (includeing Fe II lines) in the PEPSI 2018 dataset. In contrast, we measure very extreme winds on the order of 8-10 km $\mathrm{s^{-1}}$ with the same data. Like \citealt{Hoeijmakers2019}, \citealt{Cauley2019} adopt the mid-transit timing ephemeris reported in \citealt{Gaudi2017}, which differs by 5.7 minutes from our ephemeris when both are propagated to the PEPSI 2018 dataset epoch. This yields a 4.1 km $\mathrm{s^{-1}}$ blueshift in our measurement of day-to-nightside winds relative to \citealt{Cauley2019}'s. Furthermore, \citealt{Cauley2019} neglected to convert the observation timings from the $\rm{JD_{UTC}}$ timing system to $\rm{BJD_{TDB}}$, which is the timing system in which the \citealt{Gaudi2017} ephemeris is given. This results in an additional 4.4 minute difference, which corresponds to an additional 3.1 km $\mathrm{s^{-1}}$ discrepancy in our measurements. The residual difference in our measurements is $\sim$1-3 km $\mathrm{s^{-1}}$. This difference captures the discrepancy between our derived systemic RV ($v_{\mathrm{sys, PEPSI}} = -17.86\pm0.044$ km $\mathrm{s^{-1}}$ and the value adopted by \citealt{Cauley2019} ($v_{\mathrm{sys}} = -20.567\pm0.1$ km $\mathrm{s^{-1}}$ from \citealt{Gaudi2017}) as well as observational errors. Our reanalysis of previously presented datasets emphasizes the importance of precision timing in constraining the atmospheric dynamics of close-orbit planets.

\par Our revised mid-transit timing ephemeris is more reliable than the value presented in \citealt{Gaudi2017}. As described in Section 3.1.3 of  \citealt{PaiAsnodkar2021}, we refit the KELT-9 system with \texttt{EXOFASTv2} \citep{Eastman2019} to include recent TESS observations \citep{TESS2015} in addition to the follow-up lightcurves and TRES RVs from the discovery paper. The revision is especially critical for this analysis since the observations span multiple years; the error propagation using the original ephemeris can be as large as 4.8 minutes for the PEPSI 2019 dataset. Incorporating the TESS 2019 data lightcurves to span a broader temporal baseline improves the precision of the ephemeris and mitigates the issue of "stale" ephemerides. 

\par \citealt{Wong2020} reports no significant temporal variability in KELT-9 b's dayside hotspot from TESS lightcurves, which seems to at odds with our observed variability of KELT-9 b's atmospheric dynamics. However, we note that the two observations probe different pressure levels and may not be relevant in the context of variability. The phase curve observations probe $\sim$1 bar level whereas our Fe II line observations probe pressure levels at $\lesssim$1 $\mathrm{\mu}$bar.

\subsection{Impact of Transit Fitting and Orbital Dynamics on Mid-Transit Timing}

\begin{figure*}[t]
    \centering
    \includegraphics[width=0.7\textwidth]{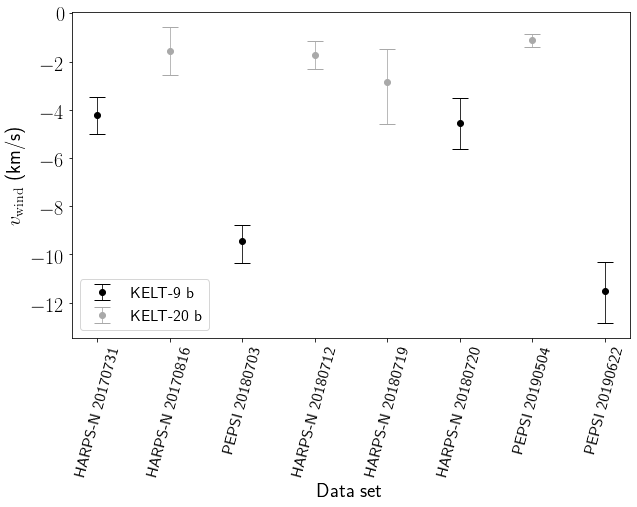}
    \caption{Day-to-nightside winds of KELT-9 b and KELT-20 b ordered chronologically along the x-axis.}
    \label{fig:UHJwinds}
\end{figure*}

One nuance to address is that our global fit does not account for the distinctly observable gravity-darkening signature in the TESS transit lightcurves. However, \citealt{Ahlers2020} similarly achieve sub-minute precision on the mid-transit timing. They show that the mid-transit timing ephemeris obtained from transit models with and without gravity-darkening differ by $\sim$0.2 minutes. Upon personal correspondence with the first author, we confirmed that they did not place a prior on mid-transit timing in their model fits (J. Ahlers, personal correspondence, August 22, 2021). Therefore, the mid-transit timings of both models (with and without gravity-darkening) inherently agree well within 1$\sigma$, especially when propagated to the different epochs of our observations. The difference of 0.2 minutes is not significant enough to yield an appreciable change in the measured day-to-nightside wind velocity as implied by Figure \ref{fig:KELT9_ephemerisWindShift}. Thus, we conclude that our mid-transit timing maintains fidelity despite neglecting the photometric effects of gravity-darkening in our global fit.

\par Other nuances that may affect the day-to-nightside wind measurement are potential eccentricity in KELT-9 b's orbit and nodal precession. Both of these effects yield a discrepancy between our modelled vs. the true orbital motion of the planet. First, we will address eccentricity. 

\par To extract day-to-nightside winds, we model the RV of the fully in-transit atmospheric absorption signature as a circular Keplerian orbit with an offset due to day-to-nightside winds (we remove the RV shift due to systemic velocity before performing this analysis). Over the small range of phases exposed during a transit, the RV signal is approximately linear over time. In principle, this offset may not be entirely due to day-to-nightside winds if KELT-9 b's true orbit were sufficiently eccentric with an argument of periapsis significantly offset from the mid-transit point along our line-of-sight. From the primary and secondary eclipse timings given by the eccentric models (models 5 and 8) in \citealt{Gaudi2017}, we constrain $|e\cos{\omega}| = 0.00170 \pm 0.00181$ using equation 3 of \citealt{Charbonneau2005}; note that this uncertainty may be underestimated as it does not account for the covariance between the primary and secondary eclipse timings. When including this constraint, we find that introducing eccentricity only changes the slope of the RV signal over the course of the transit and does not change the linearity or offset of the signal for reasonably small eccentricities ($0.001 < e < 0.2$). Since the slope of the signal ($K_{\mathrm{p}}$) is a free parameter in our fit, we conclude that our model sufficiently captures the orbital motion of the planet and the offset exclusively represents the day-to-nightside winds on KELT-9 b.

\par Nodal precession also modifies the planet's orbit in such a way that it could lead to a misinterpretation of our observations. Nodal precession can occur to a planet around an oblate star, where a non-zero gravitational quadrupole field causes the planet's orbital plane to precess. Consequently, the chord across the stellar disk transited by the planet changes over the years, which changes the orbital phases over which the planet's atmosphere is illuminated. Thus, our modelled orbital motion of the planet may not accurately capture the true orbital motion without accounting for nodal precession. The observed parameters modified by nodal precession are impact parameter (directly related to orbital inclination by the transit observable $\frac{a}{R_\star}$) and spin-orbit angle, both of which can be measured by Doppler Tomography. We estimate the effects of nodal precession on the orbital motion of KELT-9 b using preliminary measurements of these parameters across all four datasets \citep{WangInPrep}. We find the greatest difference in mid-transit timing due to nodal precession is between the PEPSI 2019 and HARPS-N 2017 datasets with a difference of 0.073 minutes. This is not significant enough to yield a measurable difference in wind velocity according to our prior precision timing argument. Furthermore, the transit timing and duration do not change drastically enough to yield a measurably significant difference in the observable orbital RVs of the planet during the course of the 4 transits. Thus, we conclude that nodal precession does not change the orbital motion of KELT-9 b enough to change the observed orbital RVs of the planet over the course of the 4 transits studied in this work. Additionally, the timescale of nodal precession ($\sim$thousands of years) is too long to induce variability in mid-transit timing and likewise the (inaccurate) measurement of day-to-nightside winds.

\subsection{Comparison with KELT-20 b}

\par To emphasize how variability of atmospheric dynamics is unique to KELT-9 b and not an artifact of our analysis from two different instruments, we apply the same procedure to PEPSI and HARPS-N observations of another UHJ, KELT-20 b. From the PEPSI line-by-line analysis, we obtain a representative global blueshifted wind speed of $-2.55^{+0.99}_{-0.76}$ km $\mathrm{s^{-1}}$, which is consistent with the previous literature value of $-2.8 \pm 0.8$ km $\mathrm{s^{-1}}$ (also traced by Fe II) in \citealt{Casasayas-Barris2019}. Cross-correlation of the PEPSI 20190504 dataset as well as the HARPS-N 20170816, 20180712, and 20180719 datasets also yield consistent results as demonstrated in the right panel of Figure \ref{fig:vwind}. The lack of supersonic winds or wind variability from our analysis of Fe II absorption in KELT-20 b's atmosphere highlights the exceptional nature of atmospheric dynamics on KELT-9 b.

\par We note that day-to-nightside wind measurements of KELT-20 b are significantly less sensitive to mid-transit timing than KELT-9 b. From our Fe II line-by-line analysis of the PEPSI dataset, we measure KELT-20 b's orbital velocity to be $168.8^{+21.0}_{-12.4}$ km $\mathrm{s^{-1}}$; this is consistent with the measurement given in \citealt{Casasayas-Barris2019}, which reports $K_{\mathrm{p}} = 174.4 \pm 14.0$ as traced by Fe II lines. Consequently, due to KELT-20 b's slower orbital velocity and larger orbital period, even an offset in mid-transit time by 10 minutes would only correspond to a 2.1 km $\mathrm{s^{-1}}$ shift in the recovered day-to-nightside wind velocity. Furthermore, a revised mid-transit timing would not induce variability in wind measurements across epochs (assuming the orbital period from literature is robust).

\par Figure \ref{fig:UHJwinds} displays representative measurements of day-to-nightside winds for KELT-9 b and KELT-20 b over time. This figure further demonstrates the dramatic variability of KELT-9 b's winds in contrast to KELT-20 b's temporal stability.

\subsection{Implications}

\subsubsection{Super-sonic winds}
For physical context, we estimate the sound speed in KELT-9 b's atmosphere by assuming a hydrogen-dominated atmosphere ($\rho = m_{\rm{H}} n$) that acts as an ideal gas ($P = n k T$) and exhibits ideal, isothermal flow ($P = c_{\rm{s}}^2 \rho$). This yields the following relation for sound speed:
\begin{equation}
    c_{\rm{s}} = \sqrt{\frac{k T}{m_{\rm{H}}}}
\end{equation}
where $c_{\rm{s}}$ is the sound speed, $k$ is the Boltzmann constant, $m_{\rm{H}}$ is the mass of a hydrogen atom, and $T$ is the temperature of the region in the atmosphere of interest. We estimate the sound speed ranges between 6.8 to 7.5 km $\mathrm{s^{-1}}$ in the region of the atmosphere probed by Fe II absorption features. In obtaining these limits, we adopt lower and upper bounds on the temperature in the region of KELT-9 b's atmosphere ranging between 1 millibar to 1 \textmu bar, the pressure range relevant to transmission spectroscopy \citep{Kirk2019}, based on the P-T profile of KELT-9 b's atmosphere presented in \citealt{Lothringer2018} (Figure 14). Thus we conclude the high altitude day-to-nightside winds we measure span all the way from sub- through supersonic speeds across epochs. While supersonic winds may be able to generate shocks locally, it is unknown to what extent these shocks can dissipate wind structures. In canonical HJs, acoustic shocks are predicted to yield $\sim$1 km $\mathrm{s^{-1}}$ variability of equatorial jets on short timescales comparable to those we observe \citep{Fromang2016, Menou2020}; this is insufficient to fully disrupt these jets but also not comparable in magnitude to our observed day-to-nightside wind variability on KELT-9 b. The measured high-speed winds are indirectly supported by the study of KELT-9 b's heat transport in \citealt{Mansfield2020}, which found that $\sim$10 km $\mathrm{s^{-1}}$ day-to-nightside winds are necessary to explain the temperature contrast observed from the phase curve of KELT-9 b (X. Tan, personal correspondence, June 15, 2021).

\subsubsection{Nightside condensation of Fe II}
\par The atmospheric absorption tracks display visibly discernible in-transit variability; see Figure \ref{fig:FeIIlines}. To quantify this variability, we stack the transmission spectra from the first and second half of the transit separately in the planet's rest-frame. We do not include observations in which the Doppler shadow overlaps with the absorption feature. Then we fit the absorption feature from each half of the transit with a Gaussian and adopt the amplitude of the Gaussian as the absorption strength. For comparison, we take the difference of the absorption strengths from the first and second half of the transit. Based on the highest SNR lines (Fe II $\lambda$4923.9, 5018.4, 5169), the second half of the transit displays stronger absorption in the PEPSI 2018 data by a mean of 1466 ppm. This is consistent with the original findings presented in \citealt{Cauley2019}. 

\par The in-transit variability of the atmospheric absorption signature seen in the PEPSI 2018 data is consistent with a picture of nightside condensation of Fe II on KELT-9 b as seen with Fe I on WASP-76 b \citep{Ehrenreich2020, Kesseli2021}. Over the course of a tidally-locked planet's transit, more of the dayside is visible on the leading limb during the first half and on the trailing limb during the second half (see Figure 3 in \citealt{Ehrenreich2020}). Assuming the same physical picture as presented in \citealt{Ehrenreich2020}: 1) Fe II condenses out on the nightside and 2) more Fe II is present in the gaseous phase at the evening side than the morning side due to a longitudinal offset of the hotspot towards the evening terminator. Invoking this picture of a chemical gradient across the planet implies stronger absorption during the second half of the transit (as empirically observed) when more dayside near the evening terminator of the planet is exposed to the observer. Alternatively \citealt{Savel2021} has shown that rather than iron rainout, the asymmetric Fe absorption in WASP-76 b is better explained by $\mathrm{Al_2O_3}$, Fe, or $\mathrm{Mg_2SiO_4}$ clouds on the cooler leading limb, which could also explain the stronger Fe II absorption we observe in the second half of KELT-9 b's transit. \citealt{Wardenier2021} also reproduces the feature with a temperature asymmetry between the leading and trailing limbs of the planet without invoking iron condensation. Stronger absorption during the second half of the transit is not quantitatively evident in the PEPSI 2019 data.

\subsection{Expectations from GCMs}
The magnitude and potential variability of day-to-nightside winds on KELT-9 b provide critical empirical context for general circulation models of UHJ atmospheres. Observational measurements of day-to-nightside winds provide constraints on the drag time constant of horizontal atmospheric circulation. Currently hydrodynamic HJ GCMs predict winds between $\sim$1-4 km $\mathrm{s^{-1}}$ and an amplitude of variability on the order of $\sim$2 km $\mathrm{s^{-1}}$ at most (if at all) over weekly timescales \citep{Showman2020, Komacek2020}. Our measurement may be significantly higher than theory in part because UHJs (especially KELT-9 b) can have a stronger day-night temperature contrast than standard HJs which drives faster winds; UHJ GCMs typically do not explicitly state wind speeds due to additional unknowns from second order effects. Additionally, these models typically do not investigate the higher altitudes of the atmosphere probed by high-resolution transmission spectroscopy and wind speeds generally increase with altitude. Our empirical work motivates future theoretical work in this extreme regime.

\par Modelling the UHJ regime is a greater challenge than HJ GCMs for several reasons. Firstly, the opacity of species at the extreme pressure-temperature conditions of UHJ atmospheres are difficult to quantify, both empirically and theoretically. Furthermore, chemistry plays a larger role in heat transport across UHJs since on the dayside, they satisfy the conditions for molecular dissociation, most notably the dissociation of the dominant species, molecular hydrogen. The current picture of UHJ chemistry predicts cooling from hydrogen dissociation on the dayside and heating from hydrogen recombination on the nightside. As a result, models that include molecular dissociation yield a weaker temperature contrast and consequently weaker day-to-nightside winds than models without it \citep{Tan2019, Roth2021}. This additional mechanism is a computationally-intensive feature exclusive to UHJ GCMs.

\par The greatest challenge to UHJ atmospheric circulation theory is incorporating magnetism in a self-consistent fashion. Purely hydrodynamic models parameterize magnetic effects as a drag term (because winds that cross magnetic field lines are dampened or potentially even fully disrupted), which may not produce accurate atmospheric circulation patterns \citep{Rauscher2013, Komacek2016, Tan2019}. On the other end of the spectrum, current magnetohydrodynamical circulation models \citep{Rogers2014, Rogers2017} do not include a radiative transfer scheme that would allow these models to be compared with observations. They also adopt the anelastic approximation, which assumes small vertical density and temperature variations and thus underpredicts wind velocities on hot Jupiters. \citealt{Beltz2021b} strikes a balance between both methodologies and adopts a spatially-varying magnetic drag prescription to account for the temperature-dependence and directionality of magnetic drag. However, this scheme assumes a static magnetic field and thus does not account for its evolution due to circulation in a fully self-consistent manner the way a MHD simulation would.

\par No UHJ GCM to-date fully captures all the physical mechanisms at play. The large day-night temperature contrast on UHJs is expected to drive winds stronger than those on HJs, but the addition of both molecular dissociation and magnetic drag in UHJs are anticipated to weaken wind velocities; this could be in tension with the rapid, trans- to supersonic wind speeds we have presented. However, magnetic drag may not affect day-to-nightside winds propagated across the poles of a UHJ as strongly as equatorial jets due to the directionality of Lorentz forces \citep{Beltz2021b}. Preliminary efforts modeling the hot upper atmospheres of UHJs, the extreme pressure-temperature regime we probe with high-resolution transmission spectroscopy, show that this holds even at the $\sim$0.1 mbar level even though the effects of magnetism are grow stronger with temperature. 

\par An added complication for KELT-9 b is that it is on a near-polar orbit around a rapidly-rotating oblate star. Thus it experiences variable heating over the course of its orbit as it passes a more intense radiation environment at the poles of the star than at the equator. This could drive the observed variability in wind dynamics in addition to the standard gyres, vortices, and other hydrodynamic instabilities predicted by hydrodynamic GCMs \citep{Fromang2016, Tan2019}. MHD simulations also predict variability of winds, especially of zonal winds in the hottest HJs which can entirely reverse from eastward to westward \citep{Rogers2014, Hindle2021}. Our measured scale of variability motivates further exploration of these models and their predictions for the day-to-nightside winds typically observed with high-resolution transmission spectroscopy in addition to the zonal winds commonly studied by theorists. 

\section{Conclusion} \label{sec:conclusion}
We revisit the atmospheric dynamics of KELT-9 b with a revised ephemeris and a new high-resolution transmission spectroscopy dataset from the PEPSI spectrograph on the LBT. Contrary to previous literature \citep{Yan2018, Yan2019, Cauley2019, Hoeijmakers2019}, we detect significant $\sim$4-10 km $\mathrm{s^{-1}}$ day-to-nightside winds as traced by line-by-line analysis as well as cross-correlation of Fe II lines (\S \ref{sec:results}). We show that the velocity shift due to the planet's orbital motion depending on the choice of mid-transit timing accounts for the disagreement with previous studies of the datasets explored in this work (\S \ref{sec:discussion}). Our work motivates future multi-epoch observations and analysis of UHJ atmospheric dynamics. The magnitude and variability of winds we observe pose new challenges for GCMs of UHJs, which require further theoretical exploration of second-order mechanisms such as magnetic drag and $\rm{H_2}$ dissociation at the extreme pressure-temperature regime probed in this work. 

\acknowledgments
A.P.A would like to thank the David G. Price Fellowship in Astronomical Instrumentation for funding her work this year. Work by B.S.G. and J.W. was partially supported by the Thomas Jefferson Chair for Space Exploration endowment from the Ohio State University.  This work is based on observations made with the Large Binocular Telescope. The LBT is an international collaboration among institutions in the United States, Italy and Germany. LBT Corporation partners are: The University of Arizona on behalf of the Arizona Board of Regents; Istituto Nazionale di Astrofisica, Italy; LBT Beteiligungsgesellschaft, Germany, representing the Max-Planck Society, The Leibniz Institute for Astrophysics Potsdam, and Heidelberg University; The Ohio State University, representing OSU, University of Notre Dame, University of Minnesota and University of Virginia. This paper includes data collected by the TESS mission. Funding for the TESS mission is provided by the NASA's Science Mission Directorate. We thank Dr. Marshall Johnson, Dr. Francesco Borsa, Dr. Jens Hoeijmakers, Dr. Xavier Dumusque, Dr. Xianyu Tan, and Dr. Johnathon Ahlers for contributing their valuable expertise regarding high-resolution spectroscopy, the HARPS-N pipeline, and the KELT-9 system. Additionally, we thank Prof. Emily Rauscher and Dr. Xianyu Tan for providing critical feedback on the manuscript and sharing their indispensable perspectives as theorists.\\ \\

\facilities{LBT (PEPSI), TNG (HARPS-N), Fred L. Whipple Observatory (TRES)} \\

\software{scipy \citep{scipy2020}, petitRADTRANS \citep{Molliere2019}, SME \citep{Valenti1996, Valenti2012}, Time Utilities \citep{Eastman2012}, emcee \citep{Foreman-Mackey2013}, george \citep{Ambikasaran2014}, EXOFASTv2 \citep{Eastman2019}}

\clearpage
\bibliographystyle{aasjournal}
\bibliography{references}

\end{CJK*}

\end{document}